\documentclass[12pt,preprint]{aastex}

\renewcommand\>{{\rangle}}
\renewcommand\bv{{\bf v}}
\newcommand\degree{{\rm o}}

\newcommand\br{{\bf r}}

\newcommand\blam{{\mbox{\boldmath $\lambda$}}}

\newcommand\<{{\langle}}

\def\cm{{\rm\,cm}}
\def\sec{{\rm\,s}}

\def\K{{\rm\,K}}

\shortauthors{Watson et al.} 
\shorttitle{MHD Masers}

\begin{document}

\title{Anisotropy of Magnetohydrodynamic Turbulence and the Polarized 
Spectra of OH Masers}
\author{William D. Watson$^{1}$, Dmitri S. Wiebe$^{1,2}$, Jonathan C. 
McKinney$^{1}$ , and Charles F. Gammie$^{1,3}$}

\affil{University of Illinois, 1110 West Green Street, Urbana, IL 61801}

\altaffiltext{1}{Department of Physics.} 
\altaffiltext{2}{Permanent
address:Institute of Astronomy of the RAS, 48, Pyatnitskaya str., 109017
Moscow, Russia.} 
\altaffiltext{3}{Department of Astronomy.}

\begin{abstract}

We consider astrophysical maser radiation that is created in the presence of
mildly supersonic, magnetohydrodynamic (MHD) turbulence. The focus is on the
OH masers for which the magnetic field is strong enough that the separations
of the Zeeman components are greater than the spectral linebreadths. A
longstanding puzzle has been the absence of the Zeeman $\pi $ components and
the high circular polarization in the observed spectra of these masers. We
first argue that the elongation of eddies along the field that has recently been recognized in MHD turbulence 
will enhance the optical depth parallel to the magnetic field in comparison
with that perpendicular to the magnetic field. We then simulate maser
emission with a numerical model of MHD turbulence to demonstrate
quantitatively how the intensities of the linearly polarized $\pi $
components are suppressed and the intensities of the nearly circularly
polarized $\sigma $ components are enhanced. This effect is also generic in the sense
that most spectral lines in MHD turbulence with Mach number $M\sim 1$ should have larger
optical depth parallel to the magnetic field than perpendicular. The effect
is reduced considerably when $M<1$. The
simulations also demonstrate that the velocity and magnetic field variations
due to the turbulence can (but do not necessarily) cause one of the  $\sigma $
components to be much more intense than the other, as is often observed for
mainline OH masers.
\end{abstract}

\keywords{masers: OH---polarization---magnetic fields---MHD---turbulence}

\section{Introduction}

Magnetohydrodynamic (MHD) turbulence is highly anisotropic (\citealt{gs95},
hereafter GS; see also \citealt{mt81,hig84}; for numerical confirmation see %
\citealt{cv00} and \citealt{mg01}). According to the GS theory, turbulent
eddies are elongated parallel to field lines, and this anisotropy increases
as one progresses to smaller and smaller scales. In traversing a turbulent
MHD gas, spectral line radiation will then typically sample regions that are
extended parallel to the field. That is, the optical depth for spectral
lines in MHD turbulence will be larger parallel to the field than
perpendicular to the field. The GS theory predicts anisotropic scattering of radio waves, consistent with observations (e.g. \citealt{nac89}).
It also predicts an angle-averaged spectrum that is consistent with measurements of radio wave fluctuations in the local interstellar medium \citep{ars95}.

Radiative transitions of atoms and molecules in the presence of a magnetic
field are split into $\sigma $ and $\pi $ components according to the change
in the quantum number for angular momentum along the magnetic field $\Delta
m=\pm 1$ or $\Delta m=0$, respectively. The absorption and emission of
radiation in these transitions by an isolated atom or molecule depends upon
the angle $\theta $ between the magnetic field and the direction of
propagation of the radiation, and is proportional to $(1+\cos ^{2}\theta )$
for $\sigma $ transitions and to $2\sin ^{2}\theta $ for $\pi $ transitions
(e.g., \citealt{cs70}, pg. 99). Radiation emitted in $\sigma $ transitions
is elliptically polarized---ranging from entirely circular at $\theta =0$ to
entirely linear at $\theta =90^{\mathrm{o}}$, whereas that due to $\pi $
transitions is 100\% linearly polarized at all angles. Hence, if the
magnitudes of the optical depths in directions close to that of the magnetic
field are larger than in other directions, the $\sigma $ components and the
circularly polarized radiation will tend to be enhanced in comparison with
the $\pi $ components and the linearly polarized radiation. For masers, the
effect of such differences can be exponentially amplified.

Among interstellar spectral lines, only the OH masers have Zeeman splittings
that are greater than the spectral line breadths and they do, in fact,
exhibit the characteristics described above. The OH masers at 1665 and 1667
MHz (``mainline'' masers) occur widely in regions of star formation and have
been especially well studied (e.g., \citealt{gb88}; \citealt{arm00}). Except
possibly for a recent observation \citep{hut02}, the $\pi $ components have
never been detected in this maser radiation. Circular polarization dominates
and in some regions the radiation is essentially 100 \% circularly
polarized. Nevertheless, a small fraction of the radiation often is linearly
polarized. These spectral characteristics have remained a puzzle (e.g., %
\citealt{reid02}) since the initial discoveries of masers in astronomy.
Though the Zeeman splitting ordinarily is much smaller for other masing
transitions of OH, the $\pi $ components seem to be absent and circular
polarization also seems to dominate for these (\citealt{baud98} [13
GHz]; see also \citealt{baud97} [6 GHz]). Gray and Field (1995) have emphasized that the $\sigma$ and $\pi$ radiation of the OH masers will tend to be beamed parallel and perpendicular, respectively, to the magnetic field. 

A secondary polarization characteristic of the OH masers is that, commonly,
only one of the two $\sigma $ components of the 1665/1667 MHz transitions is
detected from a particular maser location in observations that are performed
at the highest possible angular  resolution, though comparable numbers of
masers with single right or single left circular polarization may be
observed from the entire cluster of masers. For other masing transitions of
OH, both $\sigma $ components more often are detected at the same location
(e.g., \citealt{mor78}; \citealt{cv95}). The tendency for one of the $\sigma 
$ components of the 1665/1667 MHz transitions to be much stronger than the
other has often been attributed to shifts in the resonant frequency caused
by gradients in the Doppler velocity and in the magnetic field that
partially cancel for one of the $\sigma $ components, but not for the other %
\citep{cook66}. Actually, a velocity gradient alone is sufficient to create
a single, sharp spectral line with a single sense of circular polarization
if the change in velocity through the maser is as large as the Zeeman
splitting and the maser is at least partially saturated (\citealt{dw86}; %
\citealt{nw90}). In either case, the observation that single components with
both senses of circular polarization occur in close proximity is more
suggestive of the randomness associated with turbulence than to large scale
gradients, which would tend to produce only one sense of the circular
polarization for masers that are located close to one another.

The calculations of this Paper explore in more detail how well the
anisotropy of the medium caused by MHD turbulence can serve as the basis for
an understanding of the polarization characteristics of the OH masers, and
conversely, whether these observed characteristics may be evidence of the
nature of MHD turbulence in the interstellar gas. Standard methods are employed for
computing a number of simulations of turbulent, compressible MHD to serve as representative ``cubes'' of velocities, magnetic fields and densities of
the masing medium. These cubes are used to represent the regions of
interstellar clouds in which clusters of OH masers are found. The coupled
radiative transfer equations for the Stokes intensities are then solved as a
function of Doppler velocity by numerical integration for a large number of
uniformly spaced rays of maser radiation that pass through the cube at
various angles. A comparison of the optical depths computed without the
effect of maser saturation for the rays parallel and perpendicular to the
direction of the average magnetic field provides a robust indication that the anisotropy of
MHD turbulence has a significant effect. In examining the computed spectra, we focus
on the brightest rays. These cover only a small fraction of the entire
surface, just as the observed masers cover only a small fraction of the
surface of the cloud in which they occur. We begin with general
considerations for both the MHD turbulence and the maser polarization (\S 2)
before giving a more detailed description of the basic equations for the
calculations (\S 3). The results of the computations are presented in \S 4.
In \S 5, we discuss further the requirements for the MHD anisotropies to be
applicable in the environments of the OH masers. We also reason in \S 5 that
Faraday rotation, which has been mentioned as a possible cause for the
absence of the $\pi$ components of OH\ masers, is an unlikely explanation.

\section{General Considerations}

\subsection{MHD Turbulence}
The statistical properties of MHD turbulence can be partially
characterized by the structure function
\begin{equation}\label{CORFUNC}
v^2_{i,\blam} = \< (v_i(\br) - v_i(\br + {\blam}))^2 \>.
\end{equation}
We will use the shorthand $v_\parallel$ to indicate the structure
function for the velocity component parallel to the magnetic field, and
$v_\perp$ to indicate the structure function for either of the two
perpendicular components. The quantity $v_\lambda^2$ means
$(v_\parallel^2 + 2 v_\perp^2)$ averaged over all directions for the
separation vector $\blam$. For Kolmogorov-like turbulence, $v_\lambda
\sim v_L (\lambda/L)^{1/3}$, where $v_L$ is the three dimensional
velocity dispersion on the outer scale $L$.

At any scale $\lambda$ it is useful to consider two characteristic
dimensionless parameters: the sonic Mach number $M_\lambda \equiv
v_\lambda /c_s$ and the Alfv\'en Mach number $M_{A,\lambda} \equiv
v_\lambda/v_A$ (we will assume that $\rho$ and $v_A$ do not scale with
$\lambda$). When $M_L \equiv M \ll 1$ turbulence is approximately incompressible;
when $M_A \ll 1$ it is strongly constrained by the presence of the mean
magnetic field, which it cannot bend. Because $v_\lambda$ is typically
an increasing function of $\lambda$, $M_\lambda$ and $M_{A,\lambda}$ are
both small at small $\lambda$. MHD turbulence is therefore
incompressible and mean field dominated on small scales. This was the
case considered by GS in their theory of strong MHD turbulence.

The OH maser lines, however, will be little influenced by turbulence
unless the Doppler shift associated with bulk motion of the gas is at
least comparable to the thermal linewidth of OH. This requires $M_L \gtrsim
1$, which implies that the turbulence is compressible. There is
currently no good theory for compressible MHD turbulence. Simulations
\citep{cl03,vos03} indicate, however, that the scalings introduced by GS
may be valid up to $M_L \sim$ a few. Encouraged by this, we will employ
GS's scalings in the estimates that follow.

We want to estimate the line-center optical depth $\tau_0$ in an unsaturated maser line. For
simplicity we will make an estimate assuming that the optical depth at Doppler velocity $v$ has
the form
\begin{equation}
\tau_v \simeq {\sigma\over{v_{th}}} \int ds\, e^{-(v-v_{turb}(s))^2/v_{th}^2}
\end{equation}
where $\sigma$ is a constant, $v_{turb}(s)$ is the component of the turbulent velocity along the line of sight, $v_{th}$ is the thermal
velocity of OH, and the
integral is taken along the line of sight. Most of the optical depth will come from a region of size
$l_{th}$ where $v_{turb}(s)$ changes by of order $v_{th}$. If $l_{th}$ is smaller
than the size of the masing cloud (i.e., $v_L > v_{th}$), then $\tau_0
\sim (\sigma/v_{th}) l_{th}$. Otherwise we simply have $\tau_0 \sim
(\sigma/v_{th}) L$. We will assume that we can estimate $l_{th}$ using
the structure function (\ref{CORFUNC}). Thus to find the line center
optical depth we need to find $l_{th}$ such that $v_{l_{th}} \sim
v_{th}$.

Perpendicular to the mean magnetic field, the GS theory gives
\begin{equation}
v_{\perp}(\lambda_\perp) \approx {v_L\over{\sqrt{3}}}
\left({\lambda_\perp\over{L}}\right)^{1/3},
\end{equation}
as in the Kolmogorov theory. Then
\begin{equation}
l_{th,\perp} = L \left( {\sqrt{3}\, v_{th}\over{v_L}} \right)^{3},
\end{equation}
where we have normalized the perpendicular velocities by assuming
isotropy at the outer scale:
\begin{equation}
v_{\parallel,L}^2 = v_{\perp,L}^2 = {v_L^2\over{3}}.
\end{equation}
A more general treatment is possible, but this simplified analysis
will serve to make our point.

To find $l_{th,\parallel}$ we must invoke a weak extension of the GS
theory, because GS theory considers incompressible Alfv\'enic
fluctuations which are necessarily perpendicular to the mean field. But
models of weakly compressible turbulence \citep{lg01}, numerical models
of compressible turbulence \citep{cl03,vos03}, and of the cascade of
slow (pseudo-Alfv\'en) waves in incompressible turbulence \citep{clv02}
suggest that $v_\parallel$ should have the same spectrum as $v_\perp$.

Correlations along the field can be determined from GS's relation
$\lambda_\parallel \sim \lambda_\perp^{2/3}$, which defines a
surface of constant correlation amplitude. Then
\begin{equation}
v_{\parallel}(\lambda_\parallel) \approx {v_L\over{\sqrt{3}}}
\left({\lambda_\parallel\over{L}}\right)^{1/2},
\end{equation}
and
\begin{equation}
l_{th,\parallel} = L \left( {\sqrt{3} v_{th}\over{v_L}} \right)^{2}.
\end{equation}
Again, we have assumed isotropy at the outer scale.

Then the ratio of parallel to perpendicular optical depths is
\begin{equation}
{l_{th,\parallel}\over{l_{th,\perp}}} = {v_L\over{\sqrt{3}v_{th}}},
\end{equation}
provided that $v_L > v_{th} \sqrt{3}$ (if the turbulence is subthermal
throughout the emitting region then the parallel and perpendicular
optical depths differ only slightly). Parallel optical depths should therefore
typically be larger than perpendicular optical depths for mildly
supersonic turbulence.

When will this effect be absent? The parallel optical depth can exceed
the perpendicular optical depths only if $v_L/v_{th} \gtrsim \sqrt{3}$.
Notice that $v_L/v_{th} = M_L \sqrt{m_{\rm OH}/2\mu} \approx 2.1M_L $,
where $\mu$ is the mean molecular weight of the gas and $m_{\rm OH}$ is the mass of
the OH molecule, so only $M_L \sim 1$ is required for an interesting
ratio of optical depths parallel and perpendicular to the field, and
conversely the effect should vanish for subsonic turbulence.

The key is, of course, the anisotropy of the turbulence. We expect that
MHD turbulence is nearly isotropic unless $M_{A,l_{th}}\lesssim 1$.
That is, the field needs to be strong enough to place a dynamical
constraint on the masing eddies. Since
\begin{equation}
M_{A,l_{th}} = {v_{l_{th}}\over{v_A}} = {v_{th}\over{v_A}} \sim \beta^{1/2},
\end{equation}
the effect is absent unless $\beta \lesssim 1$.

The estimates in this section suggest that the optical depth to OH maser
lines should be larger parallel to the mean magnetic field than
perpendicular to it, provided that the turbulence is mildly supersonic
and $\beta \lesssim 1$. Our estimates are encouraging but not
conclusive because they rely on an extension of the GS theory. In
addition, strong maser spots are rare events whose frequency may not be
well described by the two-point statistics considered here. A test of
these estimates against a simulation of compressible 3D MHD turbulence
is required.

\subsection{Polarized Maser Radiation}

The transfer equations for polarized radiation (see \S 3) can readily
be solved for a uniform medium in the regime where the maser is unsaturated
and the Zeeman splitting of the components is much greater than the spectral
linebreadth so that overlap can be ignored (\citealt{gkk73a},
hereafter GKK, pg. 124). For an optical depth of magnitude $\tau _{0},$
the intensities $I_{\pm \rm{ }}$ of the two $\sigma$ components and $I_{0}$
of the $\pi$ component are given by 
\begin{equation}
I_{\pm \rm{ }}\simeq (I_{c}/2)\exp [\tau _{0}(1+\cos ^{2}\theta )/2]
\end{equation}
and 
\begin{equation}
I_{0\rm{ }}\simeq  (I_{c}/2)\exp [\tau _{0}\sin ^{2}\theta ]
\end{equation}
for propagation at an angle $\theta $ from the magnetic field when the
amplification is large as is the case for astrophysical masers. Here, $I_{c}$
is the unpolarized continuum radiation that is incident on the far side of
the maser and serves as the ``seed'' radiation. In terms of the Stokes
parameters $V$ and $Q$ that measure the circular and linear polarization,
respectively, the fractional polarizations are 
\begin{equation}
{V_{\pm }\over{I_{\pm }}}=\frac{\pm 2\cos \theta }{1+\cos ^{2}\theta },
\end{equation}
\begin{equation}
{Q_{\pm }\over{I_{\pm }}}=\frac{\sin ^{2}\theta }{1+\cos ^{2}\theta },
\end{equation}
\begin{equation}
{V_{0}\over{I_{0}}}=0,
\end{equation}
and 
\begin{equation}
{Q_{0}\over{I_{0}}}=-1,
\end{equation}
where the direction of the linear polarization is parallel (perpendicular) to the direction of the magnetic field projected on the sky when $Q$ is negative (positive).   
A reasonable estimate for the mainline OH masers is $\tau _{0}\simeq 20$. From equations (10) and (11), the intensity $I_{\pm \rm{ }}$ of the $\sigma$
components can then be seen to peak strongly at small angles $\theta $ where
circular polarization dominates. The intensity $I_{0}$ of the linearly
polarized $\pi$ components is similarly peaked in directions that are
perpendicular to the magnetic field. We would not then expect to observe the
$\sigma$ and $\pi$ components together from the same region if the magnetic field
there is uniform and the masers are unsaturated. However, statistically we
are equally likely to observe regions where the magnetic field makes large
angles with the line-of-sight and regions where it makes small angles.
Hence, in conflict with the observations, comparable numbers of $\sigma$
and $\pi$
components would be detected when a number of independent regions are
observed. The situation is even worse for understanding the absence of
$\pi$
components in the limit of highly saturated masing. All three intensities ($
I_{\pm \rm{ }}$ and $I_{0}$) become equal and independent of angle $\theta 
$ with fractional polarizations that are the same as given by equations
(12)-(15) for unsaturated masers (GKK, pg. 122).

On the other hand, if the medium is anisotropic and the optical depth
for unsaturated masing is greater in directions close to the magnetic
field than perpendicular to the field lines, then $I_{\pm \rm{ }}$ near
$\theta =0$ can be much greater than $I_{0}$ near $\theta =90^{\rm{o}}$
(or at any other angle) even for modest fractional differences in the
optical depths according to equations (10) and (11). It is then possible
that the $\sigma$ components near $\theta =0$ are strong enough to be
detected while the $\pi$ components near $\theta=90^{\rm{o}}$ or at any other angle are too weak to be detected. As
described above, the anisotropy of MHD turbulence can cause the optical
depths along the field lines to be greater than those perpendicular to
the field lines.

Trapped infrared radiation and perhaps other processes can randomize the
thermal velocities and the populations of the magnetic substates of the
masing molecules at a rate $\gamma $. This rate can be greater than the rate
$\Gamma $ at which an excited masing state decays as a result of all
processes other than the stimulated emission by the maser radiation
which proceeds at a rate $R$ \citep{gkk73b,aw93}.  For simplicity, we
take the rate $\gamma$ to be the same for randomizing both the velocities and
substate populations. A regime $\gamma >R>\Gamma $ then exists in which
the maser is saturated in the sense that the maser radiation influences
the molecular populations and the amplification is approximately linear
(as opposed to exponential) with the length of the maser in a uniform
medium, but still behaves like an unsaturated maser in retaining angular
distributions for $\sigma$ and $\pi$ components of the form given by
equations (10) and (11). The spectral line narrowing of unsaturated maser
amplification to produce subthermal linebreadths also is preserved in
this regime. Just as for completely unsaturated masers, the velocity
distribution of the masing molecules is Maxwellian and the populations
of an energy level are equal in this regime.

\bigskip

\section{Basic Methods}

\subsection{Numerical Model}

Our analysis is based on a numerical model of decaying MHD turbulence. In
the model, we integrate the equations of compressible, ideal MHD: 
\begin{equation}
{\frac{{\partial} \rho}{{{\partial} t}}} + \mathbf{\nabla}\cdot({\rho \bv}) =
0,
\end{equation}
\begin{equation}
{\frac{{\partial} v}{{{\partial} t}}} + (\bv \cdot \mathbf{\nabla}) \bv = -{
\frac{\mathbf{\nabla} p}{{\rho}}} + {\frac{(\mathbf{B} \cdot \mathbf{\nabla}
) \mathbf{B} }{{4\pi\rho}}} - {\frac{\mathbf{\nabla} B^2 }{{8\pi\rho}}},
\end{equation}
\begin{equation}
{\frac{{\partial} \mathbf{B}}{{{\partial} t}}} = \mathbf{\nabla} \times (\bv 
\times \mathbf{B}),
\end{equation}
and as usual $\rho \equiv$ density, $\bv \equiv$ velocity, $p \equiv$
pressure, and $\mathbf{B} \equiv$ magnetic field.  We use an isothermal
equation of state $p = c_s^2 \rho$, with $c_s = constant$. This equation
of state is valid in regions where the cooling time is much shorter than
a dynamical time. No forcing term is employed, and the model is
nonself-gravitating.  The scheme contains no explicit resistivity or
viscosity, although a nonlinear artificial viscosity is used to capture
shocks.  

The basic equations are integrated in a cubic, periodic domain of linear
size $L$. The initial state consists of a uniform medium, $\rho = \rho_0 =
constant$, $\mathbf{B} = B_{x0} \hat{\mathbf{x}}$, with a superposed velocity
perturbation $\delta \bv$. The $\delta \bv$ is a Gaussian, random field with
the following properties: (1) the spectrum $\< \delta \bv_{\mathbf{k}}^2 \>
\sim k^{-11/3}$ for $k_{max} > k > k_{min}$(this is a Kolmogorov
spectrum); (2) $k_{max} = 32 (2\pi/L)$; (3) $k_{min} = 2 (2\pi/L)$, (4) $
\mathbf{\nabla} \cdot \delta \bv = 0$; (5) the modes with $k = k_{min}$ had
their amplitudes set equal to the expectation value.

The power-law spectrum was adopted to hasten the relaxation of the turbulent
state. A Kolmogorov slope was chosen (rather than the $-4$ expected of
strongly compressible turbulence) because our models have a Mach number of a
few and are therefore only weakly compressible. Experience suggests that the
evolution is insensitive to the precise choice of index as long as most of
the power is concentrated in the long wavelength spatial modes.

A fixed (rather than resolution-dependent) $k_{max}$ was introduced to
permit evolutions of identical realizations of the power spectrum at
different resolutions. The $k_{min}$ was set at a wavenumber corresponding
to a wavelength $L/2$ rather than $L$ in order to increase the number of modes that
contribute significantly to the power. This reduced run-to-run variations in
the evolution, and was also the motivation for fixing the amplitude of modes
with $k = k_{min}$. The initial velocity field was made incompressible to
facilitate comparison with earlier runs (e.g. \citealt{sog98}) which used a
similar constraint in order to minimize turbulent dissipation.

The initial conditions are scaled so that $L = c_s = \rho_0 = 1$. These
quantities can then be rescaled to match physical conditions in a masing
region.

Our numerical method is based on the ZEUS scheme \citep{sn92b}. ZEUS is an
operator-split, finite difference scheme on a staggered mesh. Constrained
transport \citep{eh88} is used to guarantee that $\mathbf{\nabla} \cdot 
\mathbf{B} = 0$ to machine precision. Our version of the ZEUS code was
parallelized by one of us (JCM) and run on NCSA's Platinum cluster.

Our numerical models have 3 parameters: the rms Mach number of the initial
conditions, the mean magnetic field strength (which can be parameterized in
terms of the initial $\beta^{-1} \equiv v_A^2/c_s^2 = B_{x0}^2/(4\pi\rho_0
c_s^2)$), and the numerical resolution. We have studied runs at resolution $
66^3, 128^3$ (our default resolution) and $256^3$. Our results are
insensitive to resolution. The initial rms Mach number in all of our 
computations is four. Three values for the initial $v_A/c_s$ are 
considered---1, 3,
and 10. For each of these $v_A/c_s$, the MHD equations are integrated 
for three independent choices for the ensemble of random numbers that 
specifies the initial amplitudes of the various Fourier components $\delta \bv_{\mathbf{k}}$. We 
thus obtain three sequences in time of MHD cubes
to represent the interstellar gas for each $v_A/c_s$.

\subsection{Maser Radiative Transfer for the Stokes Parameters}

The maser radiation is calculated with the simplification that each ray is
treated as an independent linear maser---a standard simplification for
astrophysical masers. Since the focus in calculating the intensities is on
only the small fraction of the rays through the computational cube that are
the brightest and represent the observed maser features, the assumption that
they can be treated as independent is reasonable since it is less
likely that they will intersect the paths of other strong rays that are propagating in
different directions. 
In any case, interaction between the rays can only have an effect when the
maser is saturated and the effect would be to strengthen the stronger maser
beam at the expense of the weaker. Since the $\sigma$ radiation tends to be the
stronger in the scenario here, the effect would be to further enhance the
$\sigma$ radiation relative to the $\pi$ radiation, and hence to strengthen the
conclusions of this investigation that MHD anisotropy tends to suppress the $\pi$ component. In addition, the present calculations are
limited to masing under conditions of modest saturation.

The calculations are performed for an angular momentum $J=1\rightarrow
0$ transition and, as is commonly done, the results are assumed to be
indicative for molecular states with higher angular momenta. For the OH
masers, it is clear that the Zeeman frequency (sometimes ``$g\Omega $'')
is much larger than the rates for stimulated emission and other
competing processes. In this regime, the radiative transfer equations
for the Stokes parameters $I,Q,U,$ and $V$ are of the same form as for
ordinary spectral line radiation and for the purposes here can be
expressed as 
\begin{equation}
\frac{d}{ds} 
\left(
\begin{array}{c}
I \\ 
Q \\ 
U \\ 
V
\end{array}
\right)
= 
\left(
\begin{array}{cccc}
A & B & 0 & C \\ 
B & A & 0 & 0 \\ 
0 & 0 & A & 0 \\ 
C & 0 & 0 & A
\end{array}
\right)
\mathbf{\times } 
\left(
\begin{array}{c}
I \\ 
Q \\ 
U \\ 
V
\end{array}
\right)
\end{equation}
from \cite{ww01} (which rephrases GKK for the relevant regime) where 
\begin{equation}
A=(1+\cos ^{2}\theta )n(f_{+}+f_{-})+2nf_{0}\sin ^{2}\theta ,
\end{equation}
\begin{equation}
B=\sin ^{2}\theta (f_{+}+f_{-}-2f_{0})n,
\end{equation}
and 
\begin{equation}
C=2n\cos \theta (f_{+}-f_{-}).
\end{equation}
Here, the $f$'s are the Maxwellian distributions for the component of the
molecular velocity along the line-of-sight and $n$ is the difference between
the normalized populations of the magnetic substates of the upper $J=1$
energy level and the normalized population of the $J=0$ lower energy level.
The $f_{\pm }$ and $f_{0}$ are evaluated at the velocities $v_{\pm }$ and $
v_{0}$ obtained from $(1-v_{\pm }/c)\omega =\omega _{R}\pm g\Omega /2$ and $
(1-v_{0}/c)\omega =\omega _{R}$ where $\omega $ is the angular frequency of
the intensities in equation (13), $\omega _{R}$ is the resonance frequency
of the masing transition for a molecule at rest and $\pm g\Omega /2$ are the
splittings of the $m=\pm 1$ substates due to the Zeeman effect. Only a
single population difference enters in equations (20)--(22) because the
calculations will be restricted to the regime of saturation ($\gamma>R$) where the populations of the
magnetic substates of an energy level are equal due to the assumed rapid
cross/velocity relaxation. The population differences $n$ are found by
solving the rate equations in this regime using the standard idealization
of ``phenomenological'' pump $\Lambda $ and loss $\Gamma $ rates 
\begin{equation}
n=\frac{\Delta \Lambda /\Gamma }{1+4(R_{+}+R_{-}+R_{0})/3\Gamma }
\end{equation}
Here $R_{+}$, $R_{-}$, and $R_{0}$ are the rates for stimulated emission
for the $m=\pm 1$ and $0$ substates (see \citealt{ww01} for detailed
expressions for $R_{+}$, $R_{-}$, and $R_{0}$), and $\Delta \Lambda $ is the
difference between the pump rates per substate into the upper and lower
energy levels. Also, note that in the above definitions $\Delta\Lambda $ does
not involve the Maxwellian velocity distributions which are included as
the $f$ 's. As a further simplification, only the values of $R_{+}$,
$R_{-}$, and $ R_{0}$ at the line centers of the $\sigma$ and $\pi$
components in the local rest frames moving with the turbulent velocities
are used in equation (23). The completely accurate procedure would
involve averaging over the absorption profile. The inaccuracy introduced
by this simplification is negligible for this investigation since the
emergent intensities near line center are adequate for the purposes
here, and since the degree of saturation will be modest (and we have
confirmed that the differences between the results with and without
saturation are small).

External continuum radiation is assumed to provide the seed radiation for
the maser in equation (19). Whether external continuum or spontaneous
emission provides the seed radiation is unimportant for the emergent
radiation. The distance $s$ is measured along the straight-line path of the
ray and the Stokes intensities in equation (13) are expressed in
dimensionless form. They are actual intensities divided by a characteristic
``saturation intensity'' $I_{s}=8\hbar \omega _{R}^{3}\Gamma /(3\pi
c^{2}A_{E}\Delta \Omega $), where $A_{E}$ is the Einstein A-value for the
transition and $\Delta \Omega $ is the solid angle into which the maser
radiation is beamed. When $I=1$ the rate for stimulated emission is
approximately equal to the decay rate $\Gamma $ for the molecular states.
The uncertain beaming angle of the maser radiation is thus incorporated into
the definition of $I_{s}$ as is customary for relating the intensity and the
mean intensity $``J"$ for the linear maser (e.g., \citealt{ww03}).

Equation (19) is expressed in the coordinate system in which the axis of
quantization for the magnetic substates is along the magnetic field. That
is, the ``z-axis'' is along the magnetic field. Stokes $Q$ and $U$ also
depend on the orientation of the coordinate system and hence upon the
direction of $\mathbf{B}$. However, in a turbulent medium the direction of $
\mathbf{B}$ changes along the path of a ray of maser radiation. It is thus
necessary to transform $Q$ and $U$ with a rotation of the coordinate system
at each step of the integration along a ray to express $Q$ and $U$ in the
coordinate system in which equation (13) is applicable. If the magnetic
field rotates by an angle $\phi $ when projected onto the plane
perpendicular the direction of propagation, $Q$ and $U$ in the coordinate
system after the rotation are related to the same quantities $Q^{\prime }$
and $U^{\prime }$, but expressed in the system before the rotation, by $
Q=Q^{\prime }\cos 2\phi +U^{\prime }\sin 2\phi $ and $U=-Q^{\prime }\sin
2\phi +U^{\prime }\cos 2\phi $ (e.g., \citealt{chandra}, pg. 34).

When turbulent velocities along the line-of-sight $v_{turb}$ are
incorporated explicitly and $g\Omega $ is expressed in terms of the magnetic
field and the magnetic moment $\mu $, the $f$'s become $f_{\pm }=F\exp
(-(v-v_{turb}\pm \mu B)^{2}/v_{th}^{2})$ and $f_{0}=F\exp
(-(v-v_{turb})^{2}/v_{th}^{2})$. In these, $v$ and $v_{th}$ are,
respectively, the Doppler velocity at which the Stokes intensities in
equation (19) are specified and the thermal velocity $
v_{th}=(2kT/m_{OH})^{1/2}$  of the OH molecules ($F$ is the normalization
constant for a Maxwellian).

To solve equation (19), it is necessary to specify (1) the radiation that is
incident on the far side of the masing cube to serve as the seed radiation,
(2) the Zeeman splitting (or more correctly $\mu $), and (3) the pumping
rate $\Delta \Lambda $.

The incident continuum radiation is assumed to be unpolarized so that the
initial values for the Stokes intensities are $Q=U=V=0$ and $I=I_{c}$. For
an assumed brightness temperature of 30 K for the external continuum seed
radiation, decay rate $\Gamma=0.03$ s$^{-1}$, and a beaming angle $\Delta \Omega = $ $5 \times 10^{-3}$ ster,  $
I_{c}=10^{-10}$ when expressed in units of $I_{s}$.

All of the computations in the Figures are obtained with this $I_{c}$,
though we have verfied that the results are insensitive to changes in $
I_{c} $ by a few powers of ten. For the Zeeman splitting, we adopt ($\mu
B_{avg}/v_{th})=9$ for the computations in the Figures.  This corresponds
to a separation of 5.4 km s$^{-1}$ between the two $\sigma$ components when the
gas temperature is 100 K. We have also performed computations with ($\mu
B_{avg}/v_{th})=4.5$ and have verified that the essential behavior of the
spectra is unchanged within this range of Zeeman splittings which is
representative of what is observed for the OH masers with distinct Zeeman
components. We choose to specify $\mu B_{avg}/v_{th}$ instead of simply $\mu 
$ so that in all of our computations, the Zeeman components will be well
separated as in the observations and the splitting will be essentially the
same for purposes of comparison.

We consider two idealizations for $\Delta \Lambda $: $\Delta \Lambda $
is equal to the density multiplied by a constant, and $\Delta \Lambda $
is simply constant throughout the computational cube. All of the results
that we present in the Figures are obtained with $\Delta \Lambda$
proportional to density. The results are generally similar when $\Delta
\Lambda =$ constant is used, and we reason that the conclusions of our
investigation are not sensitive to this idealized treatment of the
pumping. The constant that multiplies the density for the pumping is
chosen separately for each cube so that $I=1$ for the strongest ray of
the 128$^{2}$ rays that emerge parallel to the magnetic field from the
128$^{2}$ grid points on the face of the particular cube. Representative
computations also are performed when this pumping constant is chosen so
that the strongest ray has $I=0.1$ and $I=10$. The differences are not
significant.

\section{Results}

When the effects of saturation are ignored (i.e., $n$ is set to be $\Delta
\Lambda /\Gamma $ regardless of the intensity) in solving equation (13) for
the intensities, an unsaturated optical depth $\tau $ can be obtained for
the intensity $I$, 
\begin{equation}
\tau =\ln (I/I_{c})
\end{equation}
Equation (13) is integrated along the paths of each of the 128$^{2}$ rays
that emerge from the grid points on the faces of the computational cubes.
The unsaturated optical depth $\tau $ at the peak intensity is then obtained
for each ray. Representative histograms of these optical depths are given in
Figure 1 for rays that are parallel and perpendicular to the direction of
the average magnetic field in the cubes (with the normalization that $I=1$
for the brightest ray in the parallel direction for each cube). The expected
result that the parallel rays should have larger optical depths is evident
for $v_{A}/c_{s}\gtrsim 3$. A similar trend with $v_{A}/c_{s}$ is evident in
all of the 81 cubes that were studied. The 81 cubes were obtained by considering the solutions at nine equally spaced times (separated by $\Delta t=0.2L/c_{s}$) for each of the nine sets of initial conditions for which the MHD equations are integrated. Although there is
considerable overlap in the histograms for $v_{A}/c_{s}=3$, the masing spots
are represented by the brightest rays from the face of the cube and these
are due to only a small number of the largest optical depths. There is no
overlap even at $v_{A}/c_{s}=3$ for the tails of the distributions at large
optical depth. It might be surprising that the perpendicular optical depths
tend to be larger than the parallel optical depths in our calculations 
for $v_{A}/c_{s}=1$. This result is somewhat artificial. It occurs because the turbulent velocities are
smaller in the MHD computations for $v_{A}/c_{s}=1$ than for $v_{A}/c_{s}=3$
and $v_{A}/c_{s}=10$, and they also are smaller than the thermal velocity $
v_{th}$ of the OH. When $v_{turb}$ is less than $v_{th}$ in $f_{\pm }$ and $
f_{0},$ the effect of the MHD anisotropy of $v_{turb}$ to enhance the
parallel, relative to the perpendicular, rays is reduced or disappears. In
contrast, the variation in ($\mu B/v_{th})$ along a ray is still significant
(at least for the adopted $\mu B_{avg}/v_{th}=9$) and reduces the optical
depths for the $\sigma$ components, but not for the $\pi$ components since $f_{0}$
does not depend on $B$. We have verified that the histograms for the
parallel and perpendicular rays coincide for $v_{A}/c_{s}=1$ when the Zeeman
splitting is set to zero, as would be expected. 

A factor in the
decrease in the anisotropy of the optical depths in Figure 1 with decreasing $v_{A}/c_{s}$ must be the decrease in the turbulent
velocities that results when the initial $v_{A}/c_{s}$ is reduced while the initial rms turbulent velocities remain fixed in the MHD simulations. The rms values for the components of the turbulent velocities in a specific direction (i.e., parallel or perpendicular to the direction of the magnetic field) are approximately 0.9$c_{s}$, 0.7$c_{s}$ and $0.4c_{s}$ for the turbulent cubes  with initial $v_{A}/c_{s}=$10, 3, and 1, respectively, in Figures 1-4. However, this cannot be the only factor. In the turbulent cubes  from the time evolution of the MHD equations that we use, the component rms velocities for $v_{A}/c_{s}=3$ decrease to approximately
$0.4c_{s}$ without a significant change in the anisotropy as indicated by histograms such as in Figure 1. A key difference emerges when we examine the dispersions (i.e., standard deviations) in velocity along the paths of individual rays. For the simulated turbulent cubes with $v_{A}/c_{s}=3 $ and $ 10$, the average dispersion in the component of the velocity along the line of sight is much smaller for rays that propagate parallel to the magnetic field than for rays that propagate perpendicular to the magnetic field. The two dispersions are equal for turbulent cubes computed with $v_{A}/c_{s}=1$ (for all $v_{A}/c_{s}$, the rms of the components of the velocities and of the turbulent magnetic fields parallel and perpendicular to the magnetic field are equal). This indicates that the turbulent medium becomes significantly anisotropic when $\beta=c_{s}^{2}/v_{A}^{2}$ is near 0.1, at least in our calculations. 

The coupled equations (19) are now integrated with the effects of saturation
included with rapid cross/velocity relaxation as described in the foregoing
Section (i.e., $n$ is given by equation 23) for turbulent cubes that are
computed with $v_{A}/c_{s}=3.$ Computations are performed for a range of
intermediate angles as well as for propagation parallel and perpendicular to
the magnetic field. At each angle, the coupled equations are solved for the
rays that emerge from the 128$^{2}$ grid points on the face of a cube. The
peak intensity of the brightest ray at each angle is plotted in Figure 2 for
representative results. To agree with the observations, all $\pi$ components
and rays with large fractional linear polarization must be too weak to be
detected. Hence, the focus is on the brightest rays. Since there is
considerable scatter in the intensities in Figure 2, we encompass the
intensity points with a shaded band to provide a sense of the behavior of
the angular variation. The main result in Figure 2 is that the intensities
within approximately 20$^{\rm{o}}$ of the direction of the average
magnetic field are greater by a factor of about 100 or more than those at angles near 90$
^{\rm{o}}$. All of these more intense rays near 0$^{\rm{o}}$ are
$\sigma$
components as indicated their high fractional circular polarization, as well
as by their Doppler velocities which are quite close to $\pm \mu B_{avg}$.
All of the weaker rays near 90$^{\rm{o}}$ are 100 \% linearly
polarized and are $\pi$ components (they are at essentially zero Doppler
velocity). Note that, as discussed in \S 3, the pumping is chosen so
that $I=1$ for the brightest ray that is parallel to the average magnetic
field in each cube. When there is no turbulence (the cube is uniform) and
the masing is treated in the unsaturated limit, the angular variations of
the intensities for $\sigma$ and $\pi$ components are given by the simple
functions of equations (10) and (11). For comparison, these functions also are
plotted in Figure 2 with $\tau _{0}$ chosen so that $I=1$ at 0$^{\rm{o}}$,
as well ( it also follows in this case that $I=1$ at 90$^{\rm{o}}$).
Finally, not only is the polarization of the single brightest ray of a
turbulent cube the same as that of a ``pure''  $\sigma$ component at angles
less than about 20$^{\rm{o}}$ (essentially 100\% circular), but the
same is true for all of the rays at these angles that have intensities
within 1\% of the brightest rays. Likewise, the rays near
90$^{\rm{o}}$ having intensities within 1\% of the brightest rays at that
angle are 100\% linearly polarized (as expected for $\pi$ components) and
have the Doppler velocity expected for $\pi$ components. Thus, only
$\sigma$
components within about 20$^{\rm{o}}$ of the average magnetic field will be
detected from the turbulent cube in Figure 2 when the observational
sensitivity is 1\% of the intensity of the brightest maser spot.

A grayscale map of the intensities that emerge from the surface of a
representative cube with $v_{A}/c_{s}=3$ and parallel to the average
magnetic field is shown in Figure 3. The general appearance seems compatible
with that of observational maps. The spectra in the side panels of the three
adjacent rays within a cluster demonstrates that a single feature of right
circular polarization, a single feature of left circular polarization and a
Zeeman pair can be in close proximity. The spectrum of the strongest feature
in the entire map is a single feature that is 100\% circularly
polarized. All are at Doppler velocities expected for $\sigma$ components ( $
\pm \mu B_{avg}$).

Finally, a histogram is given in Figure 4 for the ratio of the peak
intensities of the weaker to the stronger feature of a Zeeman pair for all
rays with intensities greater than 3\% of the most intense ray for the
two turbulent cubes with $v_{A}/c_{s}=3$, one of which is the cube used in Figure 1. The 3\% is an estimate for
the weakest features that will be recorded in the observations. One of the
histograms in Figure 4 demonstrates that the description of the turbulence being used
here is consistent with the observational result that only one of the two
circularly polarized Zeeman components frequently is detected for the
mainline 1665/1667 MHz masers. For the variations of the velocities and
magnetic fields in other turbulent cubes (e.g., the other histogram in Figure 4), we find the full range of
possibilities---in some the Zeeman pairs and unpaired components occur with
comparable frequency, in others the Zeeman pairs are much more numerous and
finally, there are many in which unpaired components are completely absent.
The rays in the histogram are subdivided according to peak intensity to
explore whether the likelihood for an unpaired component depends on
intensity. No such dependence stands out, though the Zeeman pairs are not
among the most intense subgroup in this histogram.

The most intense rays from a cube depend on the tails of the distributions
in Figure 1 at large optical optical depths. The tails of the distributions
are, as is usually the case for tails of distributions, likely to be sensitive to details that are not
well described in idealized models. However, the differences in the
average optical depths in the distributions in Figure 1 provide compelling
and robust evidence that the $\sigma $ components will tend to be much more intense
than $\pi$ components.

\section{Discussion}

For the MHD anisotropy to be effective in altering the optical depths, the turbulent velocities must be
significant in comparison with the thermal dispersion of the OH molecular
velocities. This minimum requirement is
easily consistent with the observational data.  The thermal velocity $v_{th}$  for OH is $\sim 0.3$ km s$^{-1}$. In contrast, velocity
dispersions of at least a few km s$^{-1}$ are
typical within the region containing a cluster of maser spots (e.g.,
\citealt{arm00}). The speed of sound for a gas
of mostly molecular hydrogen at 100 K is 0.6 km s$^{-1}$.  A second condition --- that the magnetic pressure exceed
the thermal pressure ($\beta \lesssim 1$) --- also seems to be satisfied.
The observed magnetic field strengths typically are a few to 10 milligauss. Hence, $\beta =c_{\rm{s}}^{2}/v_{\rm{A}}^{2} \simeq
0.1[(N/10^{7}$cm$^{-3})($T$/100$ K$)/(B$/4mG$)^{2}]$ where a
gas density of $10^{7}$ cm$^{-3}$ H$_{\rm{2}}$ molecules is adopted as
benchmark density based on calculations for the pumping of the masers
\citep{cw91,pk96}.  Although there are
uncertainties in the number density $N$ and the kinetic temperature T, as
well as in the precise maximum value of $\beta $ that is allowed, it is
reasonable to expect from the observations of $B$ that $v_{\rm{A}
}/c_{s}\gtrsim 3$ for which our calculations in Figure 1 indicate that the
magnetic field will be strong enough to create the necessary anisotropy. 
More generally, equipartition of energy between the turbulent magnetic field
and the turbulent motions is suggested both by observations of the
interstellar medium and by MHD simulations. Then, even in the absence of a
significant uniform magnetic field, the turbulent magnetic field satisfies  $B/(4\pi\rho)^{1/2}\simeq v_{t}$, which is equivalent to the statement $\beta
\lesssim 1$ and that $V_{\rm{A}
}> c_{s}$ if the turbulent motions are supersonic.

In addition to (1) the ratio of the turbulent to the OH molecular velocity and (2) the ratio of the thermal to the magnetic pressure ($\beta$), another key consideration is the largest scale length (or minimum
wavenumber) at which the turbulence is introduced into the medium ---
the outer scale length. Since this determines the maximum length for an
elongated masing volume, injection of the turbulence at the largest
scale lengths is most effective for creating anisotropy in the optical depths. As
noted in \S 3, a minimum wavenumber $k_{\min }$ that corresponds to the
maximum wavelength $L/2$ is used for the calculations here. The relationship between the length
$L$ of the edge of the cubic volume within which the fields are computed and actual distances within interstellar clouds
can be estimated from the size of a masing spot. In the maps that we
compute such as Figure 3, the spectra and intensities change
significantly over distances of only a few grid points. We naturally
associate this dimension with the ``spot size'' $\sim 10^{14}\cm$ for
mainline OH masers so that the 128 grid points along the edge of a cube
correspond to $L\sim 10^{16}\cm$. This $L$ is closer to the dimensions
of one of the dozen or so clusters of masers in a masing cloud (e.g.,
\citealt{gb88}) than to the entire cloud.  A few computations with
$v_{\rm{A}}/c_{\rm{s}}=10$ were performed where the maximum wavelength
for injecting the turbulence was reduced to $L/5$.  The differences
between the mean optical depths parallel and perpendicular to the
magnetic field analogous to those in Figure 1 are reduced, though the
parallel optical depths can still be the larger by about 50\%
---sufficient for the maser emission along the magnetic field to
be more intense by a factor of 100 or more than that perpendicular to
$B$.

The premise of our calculatons that the OH 1665/1667 GHz masers are not highly saturated is supported by the
observation that the spectral line breadths tend to be narrower than the
thermal breadths for the kinetic temperatures that are expected in the
masing gas. The
masers may still be somewhat saturated ($R>\Gamma $) as long as the rate $\gamma$ for
velocity/cross relaxation satisfies $\gamma >R$.

The OH masers in interstellar clouds are believed to arise in gas that has been influenced by shock waves. Hence, our use of decaying turbulence (as opposed to continuously driven turbulence) seems most appropriate.

Dissipative processes will tend to erase structure in the gas at small
scales.  We need to check that the dissipation scale is smaller
than our numerical resolution in the simulations and smaller than the
maser spots $\sim 10^{14}\cm$.  For $N \sim 10^7 \cm^{-3}$, $T \sim
10^2\K$, and a fractional ionization $x \sim 10^{-6}$, the dominant dissipative
process will be ambipolar diffusion.  In linear theory, ambipolar
diffusion critically damps an Alfv\'en wave when the wave frequency $k
v_A$ (calculated in the absence of ambipolar diffusion) is comparable with
the collision frequency $\nu_{nI} \approx x N \<\sigma v\>_{In}$ between
a neutral and an ion, where $\<\sigma v\>_{In} = 1.9 \times 10^{-9}
\cm^3 \sec^{-1}$ \citep{kp69}.  This corresponds to $\lambda_D \sim
10^{13}(T/100\K)^{1/2} \beta^{-1/2} (10^7\cm^{-3}/N)(10^{-6}/x)$, or
somewhat less than the size of a maser spot.

The MHD anisotropy for optical depths parallel and perpendicular to the
magnetic field also applies to other spectral lines---both masing and
non-masing (or ``thermal'') spectral lines. These differences in the
optical depths may lead to anisotropic (or ``$m$-dependent'' ) pumping
as has been considered previously for the linear polarization of both
masing \citep{ww83,ww84} and thermal (\citealt{gk81,gk82};
\citealt{dw84}; \citealt{lis88}) spectral lines where the Zeeman
splitting is much less than the spectral linebreadth. When applied to
the pumping of OH masers, this anisotropy may also contribute
significantly to the polarization of OH masers and to the suppression of
the $\pi$ component.

In contrast to our idealized calculation with phenomenological pumping rates
and only two energy levels, \cite{gf95} calculated the excitation
and transport of polarized OH maser radiation in a magnetic field including
explicitly the magnetic substates of a number of the energy levels of the
OH molecule. The important idea that $\sigma $ radiation would
be strongly beamed in the direction parallel to the magnetic field and $\pi$
radiation beamed perpendicular to the magnetic field was recognized there.
However, at that time no cause was recognized for suppressing the
propagation of maser radiation perpendicular to the magnetic field in
comparison with that parallel to the magnetic field, and hence for ultimately suppressing
the $\pi$ components in comparison with the $\sigma $ components.

Faraday rotation within a source tends to destroy the linear
polarization of radiation and has sometimes been mentioned as a possible
cause for the absence of the linearly polarized $\pi$ components and for
the general tendency for circular polarization to dominate in OH masers.
For unsaturated masing and large Faraday rotation, equations (10) and (11)
would become (GKK, eq. 56 \& 57) \begin{equation} I_{\pm \rm{ }}=(I_{c}/2)\exp [\tau
_{0}(1+\left| \cos \theta \right|)^{2}/4] \end{equation} and
\begin{equation} I_{0\rm{ }}=(I_{c}/2)\exp [\tau _{0}\sin ^{2}\theta /2]
\end{equation} The modification is thus equivalent to reducing the magnitude of the 
optical depth of the $\pi$ component by a factor of two at all angles while
leaving the intensity of the $\sigma$ components essentially unchanged
at angles near $\theta =0$ where they are most intense$.$ Its effect on
the relative intensities of the $\pi$ and $\sigma$ components could be at
least as great as what we have calculated for the influence of MHD
turbulence. Faraday rotation depends on the column density of free
electrons through the masing region and on the component of the magnetic
field parallel to the line of sight (e.g., \citealt{spit78}, pg.  66).
While an adequate estimate of the latter can be obtained from the
observations, the column density of free electron is highly uncertain
and reasonable estimates seem to allow the possibility that Faraday
rotation could be either large or negligible for the 1665/1667 MHz OH masers.
However, a compelling argument against Faraday rotation as the general
cause for the absence of the $\pi$ components is the observation that,
although circular polarization dominates, linearly polarized radiation
frequently is detected in the $\sigma$ components. If Faraday rotation were always
strong enough to suppress the $\pi$ components, it would also be strong
enough to eliminate completely the linear polarization of the $\sigma$
components---in conflict with the observations. Again, considering
Faraday rotation in the regime of saturated masing does not help. In the
limit of large Faraday rotation and high saturation, the $\sigma$
components must be completely circularly polarized and, in principle,
the three intensities $I_{\pm \rm{ }}$ and $I_{0\rm{ } }$ are equal at
all angles (GKK, eq. 47). 

Although fractional linear polarizations of 10\% or so are probably  compatible with our calculations as they stand for suppressing the $\pi$ components, higher linear polarizations ($\gtrsim 30\%$) are detected in some clouds in a small fraction of the components and these are interpreted as $\sigma$ components (e.g., \citealt{gb88,gch02,sh02}). Note that circular polarization in a feature may not assure that the feature is a $\sigma$ component since a fraction of the linear polarization can, in principle, be converted into circular polariztion by Faraday rotation or magnetorotation. If these are in fact $\sigma$ components, they present a challenge. This radiation must, presumably, be emitted at large angles $\theta$ relative to the magnetic field as seen from equation (13) and Figure 2. At these large angles, the optical depths of the $\pi$ components would be expected to be comparable to the optical depths of the $\sigma$ components and hence to be detected in similar (albeit small) numbers---in apparent conflict with the observations.  However,we reason that there are additional considerations which may still allow the suppression of the $\pi$ components to be understood within the context of the anisotropy of MHD turbulence. (1) Our simplifications in considering only two energy levels and in ignoring any interaction between beams of maser radiation propagating in different directions may be inadequate. When these effects are included, Gray and Field (1995) find that the $\pi$ components are further suppressed if the optical depths favor directions along the field lines. (2) The anisotropy noted above in the optical depths of the radiation involved in the pumping will tend to cause the populations of the magnetic substates to be unequal, and also potentially contribute to suppressing the $\pi$ components. 
Despite this deficiency in the present calculations, it seems clear that MHD turbulence can create an anisotropic environment in interstellar clouds, and that the $\pi$ components will tend to be suppressed while the circular polarization is favored in the spectra of the mainline OH masers in this environment.

\bigskip

This work was supported in part by a NASA GSRP Fellowship Grant S01-GSRP-044
to JCM, an NCSA Faculty Fellowship for CFG, and NSF Grants AST-9988104 and
AST-0093091, and NASA grant NAG 5-9180. Some of our computations were done on NCSA's platinum cluster. We are grateful to H. W. Wyld for helpful discussions.

\clearpage

\begin{figure}
\epsscale{0.5}
\plotone{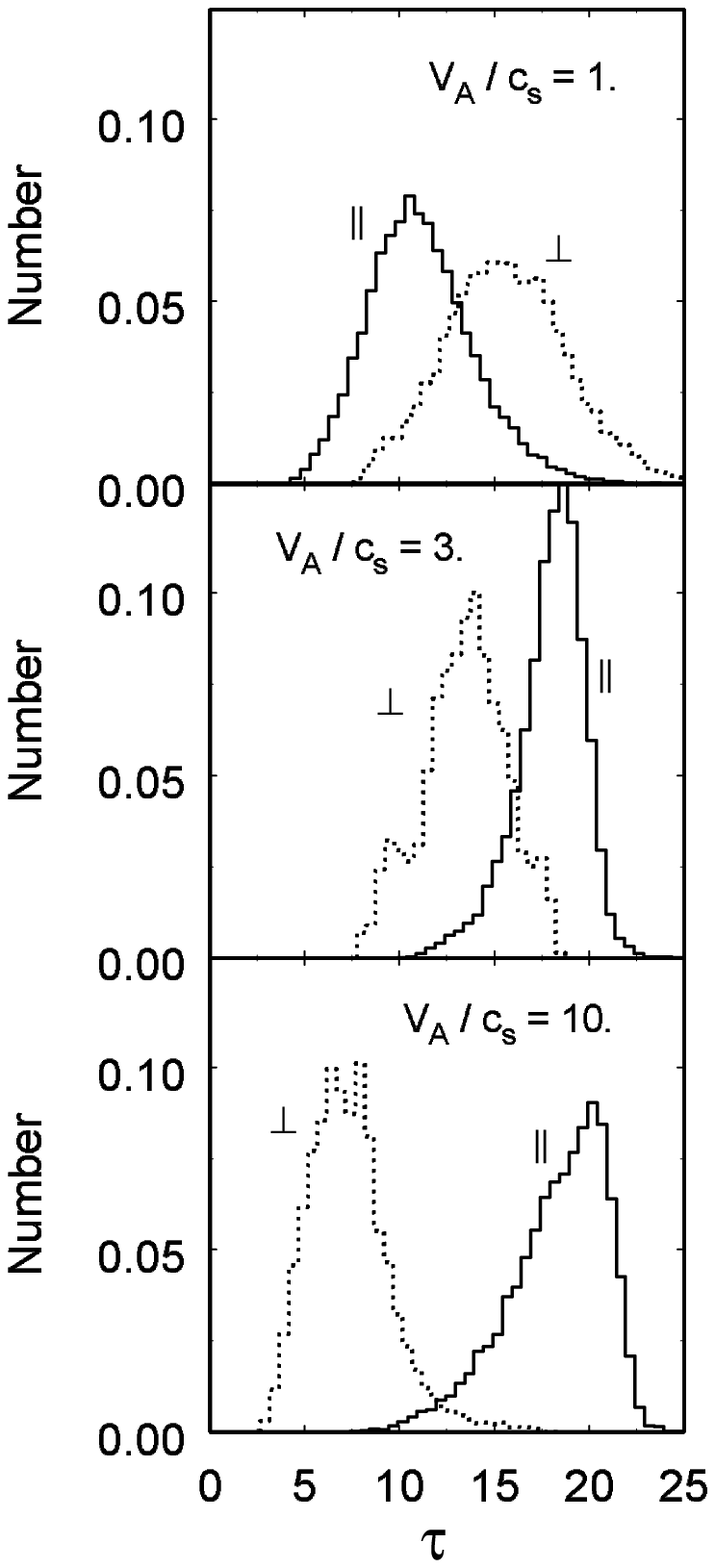}
\caption{ Representative histograms for the number of rays with unsaturated
optical depths $\protect\tau $. The optical depth is obtained at the peak
intensity of each ray. The numbers are given as a fractions of the 128$^{2}$
rays from the surface of a turbulent cube within intervals of 0.5 in $
\protect\tau $. Separate histograms (as indicated) are shown for rays that
propagate parallel and perpendicular to the average magnetic field of the
turbulent cube for the three choices $v_{A}/c_{\rm{s}}=1,3,$and 10.}
\end{figure}

\begin{figure}
\epsscale{0.9}
\plotone{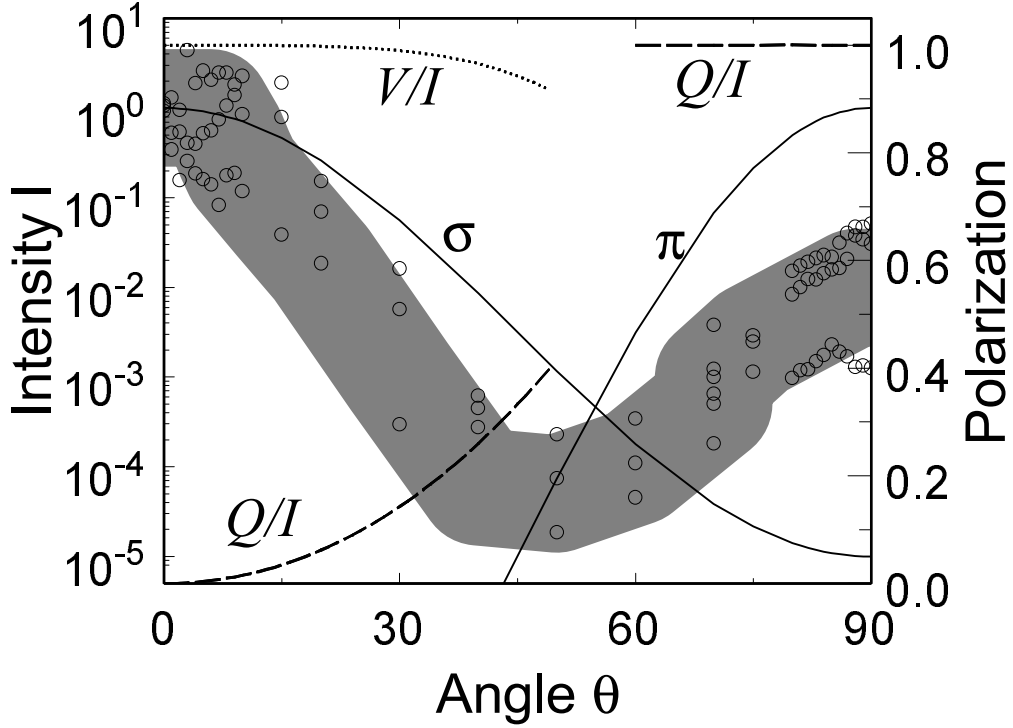}
\caption{
The intensity $I$(indicated by open circles and in units of the saturation intensity $I_{s}$) for the
brightest maser ray that emerges from a turbulent cube is plotted for
several representative cubes with $v_{A}/c_{\rm{s}}=3$ as a function of
the angle between the ray and the magnetic field, and with the constant
for the pumping chosen so that $I=1$ at 0$^{\degree}$ for each cube. The
intensities that individual $\sigma$ and $\pi$ components would have for
unsaturated masing in a uniform medium are plotted (solid lines) for comparison
(equations 10 and 11). At angles less than 55$^{\degree} $, the fractional
circular ($V/I$) and linear ($Q/I$) polarizations are plotted for this
unsaturated $\sigma$ component (equations 12 and 13); for angles greater
than 55$^{\degree}$, the fractional linear polarization is plotted for
the $\pi$ component (equation 15).  The polarizations of the intensities
of all rays at a particular angle that are computed using the turbulent
cubes and for which the intensities are within 3\% of the brightest ray
at that angle have polarizations that are essentially identical with
those of the unsaturated $\sigma$ and $\pi$ components in a uniform
medium. The fractional circular and linear polarizations of the unsaturated components in a uniform medium are indicated by dashed and dotted lines, respectively.
}
\end{figure}

\begin{figure}
\rotatebox{-90}{
\epsscale{0.7}
\plotone{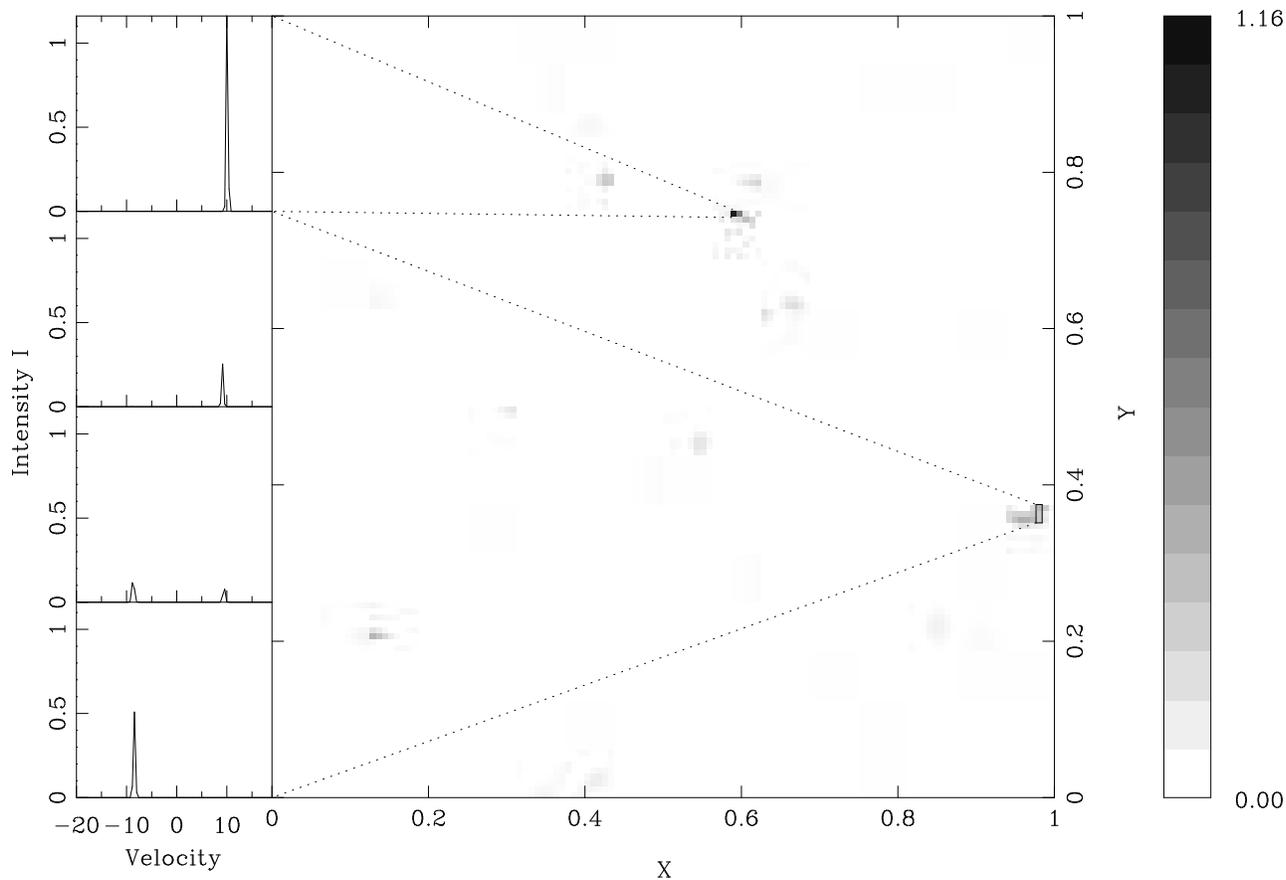}}
\caption{ Representative grayscale map of the intensities $I$(in units of $
I_{s}$) that emerge from the face of a turbulent cube with $v_{A}/c_{\rm{s}
}=3$ and parallel to the direction of the magnetic field. The panels on the
left show the spectra for the rays taken from selected locations on the map
as indicated. The Doppler velocity scale is in thermal breadths $v_{th}$ for
the OH molecule. }
\end{figure}

\begin{figure}
\epsscale{1.0}
\plotone{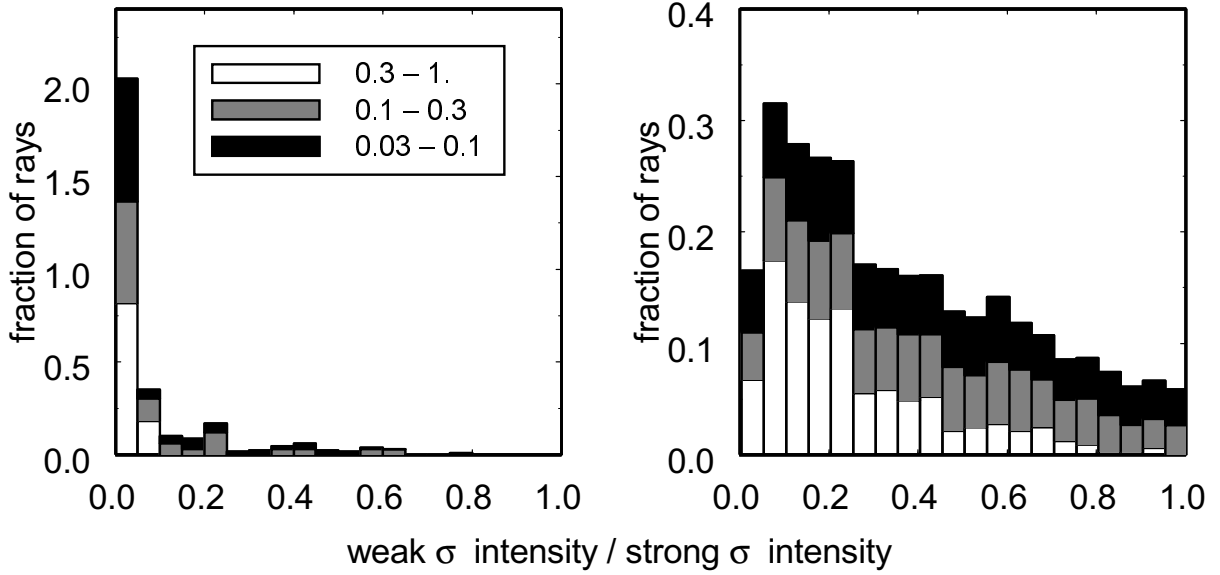}
\caption{ Examples of the ratio of the weaker to the stronger of the $\sigma$ components of
the Zeeman pairs for turbulent cubes with $v_{A}/c_{\rm{s}}=3$. The lefthand histogram is computed for the turbulent cube in
Figure 2. Only the rays with intensities greater than 0.03 of the brightest
ray are included. The rays are grouped as fractions of the intensity of the
brightest ray. Those with intensities from 0.03 to 0.1 of the brightest ray
are indicated by black; those with intensities between 0.1 and 0.3 of the
brightest, by gray; those with intensities between 0.3 and the brightest, by
white. The scale for each of these groups is normalized so that for each
group, the sum in all columns of a type is one (and the sum of all columns
is three). The bins are 0.05 in width. For some 75\% of all rays in
the lefthand histogram, the ratio is less than 0.05. The righthand histogram is an example where there are more Zeeman pairs. }
\end{figure}

\end{document}